# Big Data-Driven Fraud Detection Using Machine Learning and Real-Time Stream Processing

Chen Liu, Hengyu Tang, Zhixiao Yang, Ke Zhou, Sangwhan Cha
*Harrisburg University of Science and Technology*
Harrisburg, PA, USA
(cchen32,htang1,zyang30,kzhou, scha)@harrisburgu.edu

*Abstract—* **In the age of digital finance, detecting fraudulent transactions and money laundering is critical for financial institutions. This paper presents a scalable and efficient solution using Big Data tools and machine learning models. We utilize real-time data streaming platforms like Apache Kafka and Flink, distributed processing frameworks such as Apache Spark, and cloud storage services AWS S3 and RDS. A synthetic dataset representing real-world Anti-Money Laundering (AML) challenges is employed to build a binary classification model. Logistic Regression, Decision Tree, and Random Forest are trained and evaluated using engineered features. Our system demonstrates over 99% classification accuracy, illustrating the power of combining Big Data architectures with machine learning to tackle fraud.**

*Keywords—Fraud Detection, AWS, Anti-Money Laundering, Big Data Architecture*

## I. INTRODUCTION

The rise of fraudulent transactions and money laundering poses increasingly severe challenges to financial institutions, leading to substantial financial losses and reputational harm. With the widespread adoption of digital payments—accelerated by the COVID-19 pandemic—the urgency for rapid and accurate fraud detection has intensified. This societal shift has resulted in an exponential growth in both transaction volume and the availability of detailed metadata, offering new opportunities for data-driven detection approaches.

However, this data abundance also introduces scalability and performance challenges. While richer datasets enable more precise identification of fraudulent activity, they simultaneously demand robust and efficient analytical systems capable of handling large-scale data streams in real time.

In response to these challenges, this study leverages a synthetic transaction dataset from 2023 to explore a Big Data–based framework for fraud detection. Our approach focuses on two key dimensions: (1) extracting meaningful fraud-related insights, and (2) integrating scalable Big Data technologies to support high-throughput analysis. The proposed methodology aims to address both the accuracy and performance requirements of modern financial fraud detection systems. The remainder of this paper is organized as follows: Section II provides background information. Section III details the system implementation. Section IV presents machine learning modeling and results. Finally, Section V concludes the paper.

## II. BACKGROUND

Financial fraud, including transaction fraud and money laundering, poses a growing threat to global financial systems. In 2024 alone, consumers in the United States reported losses exceeding $12.5 billion due to fraudulent activities, a 25% increase from the previous year [1]. This increase highlights the escalating complexity and frequency of fraud targeting financial institutions and individuals.

The widespread adoption of digital payments, particularly accelerated by the COVID-19 pandemic, has significantly increased the volume and velocity of financial transactions. The global digital payment market is projected to reach $20 trillion by 2025, driven by mobile banking, online commerce, and peer-to-peer platforms [2]. This transition has introduced new opportunities for data-driven fraud detection, as each transaction now includes a wealth of metadata such as timestamps, geolocation, device identifiers, and contextual details.

However, the massive scale of modern financial data streams presents significant challenges. Traditional rule-based detection systems are insufficient in addressing the speed and sophistication of modern fraud schemes. Instead, advanced analytics and real-time anomaly detection are required to meet the demands of today's digital financial environment [3].

Recent advancements in Big Data technologies—such as Apache Spark, Kafka, and Hadoop—have enabled scalable, distributed processing of large datasets, making them well-suited for financial fraud detection systems. When integrated with machine learning techniques, these platforms can detect complex fraud patterns and adapt to evolving threats by leveraging diverse data sources [4].

This study aims to address these challenges by developing a Big Data–driven fraud detection framework using a synthetic transaction dataset from 2023. The approach focuses on two critical aspects: (1) the extraction of meaningful fraud-related insights and (2) the deployment of scalable Big Data technologies to enable real-time, high-throughput analysis.

## III. SYSTEM IMPLEMENTATION

We divide the system implementation into two key components: (1) the selection of appropriate Big Data tools, and (2) the design of the workflow architecture. For each component, we provide an in-depth analysis of the design rationale and demonstrate how the proposed system supports the effective realization of the aforementioned objectives.

## A. BIG DATA TOOL SELECTION

To efficiently process and analyze large-scale financial transaction data, the proposed system integrates a suite of Big Data technologies, each selected for its strengths in handling high-volume and high-velocity data streams.

Apache Kafka serves as the foundation for real-time data ingestion and transaction pipeline management. As a distributed event streaming platform, Kafka enables scalable, fault-tolerant stream processing through its publish–subscribe architecture. It is particularly suited for real-time event streaming and seamless data migration across system components.

Apache Flink is employed for real-time stream processing with minimal latency. Its dual capabilities for both stream and batch processing make it highly effective in analyzing high-velocity transaction data, such as those from credit card activity, where real-time decision-making is critical.

Apache Spark provides a unified engine for large-scale, distributed batch processing. In addition to supporting streaming analytics, Spark enables machine learning and graph computation at scale. Its extensive built-in libraries and high compatibility with other Big Data platforms make it an ideal core processing engine for the system.

For data storage, the system utilizes Amazon S3 and Amazon RDS to manage unstructured and structured transaction data, respectively. Amazon S3 offers scalable object storage, while Amazon RDS provides relational database support. Both services are part of the AWS ecosystem, offering secure, scalable, and cost-effective cloud-native infrastructure.

MLflow is adopted to manage the machine learning lifecycle, including model versioning, experiment tracking, and deployment. In high-traffic systems where models must be frequently updated or retrained, MLflow provides a reliable framework for orchestrating continuous model iteration and performance monitoring.

For data visualization, the system incorporates Python libraries such as Matplotlib and Seaborn, alongside business intelligence tools like Tableau and Power BI. These tools support comprehensive data analysis and effective presentation of system performance metrics and fraud detection insights.

Together, these technologies form a cohesive and scalable infrastructure capable of supporting real-time fraud detection in modern financial systems.

## B. WORKFLOW ARCHITECTURE DESIGN

The workflow architecture is structured into two distinct processing stages as shown in Figure 1: the real-time processing layer and the batch processing backend. Each stage is designed to address specific performance and analytical requirements, leveraging the previously selected Big Data tools to ensure both speed and accuracy.

The real-time processing stage focuses on the rapid ingestion, streaming, and preliminary analysis of transaction data. This layer is critical for detecting fraud as it occurs, minimizing response latency, and enabling immediate intervention. Apache Kafka serves as the event streaming backbone, facilitating real-time data ingestion and transport. Apache Flink and Apache Spark Streaming are employed to perform lightweight transformation, enrichment, and anomaly detection with minimal delay.

In contrast, the batch processing backend is responsible for deeper analytics, model training, and long-term trend analysis. This stage operates on historical transaction data accumulated over time, enabling the extraction of complex fraud patterns and the refinement of detection models. Apache Spark supports distributed batch processing and integration with MLflow for orchestrating machine learning workflows, while Amazon S3 and RDS manage the persistent storage of unstructured and structured data, respectively.

By decoupling real-time and batch processes, the proposed architecture balances the need for low-latency fraud detection with the ability to derive actionable insights through high-throughput, large-scale data analysis. This hybrid architecture enables scalable, adaptive, and continuous improvement of fraud detection capabilities in modern financial systems.

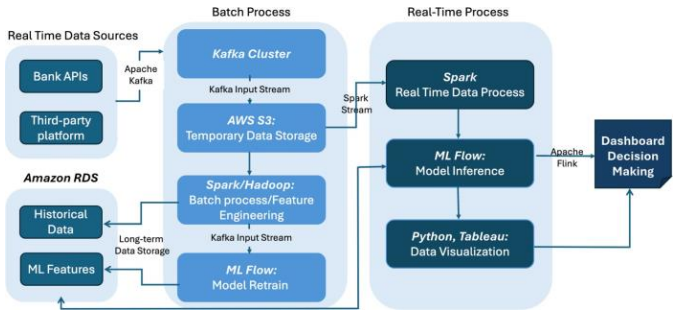

**Fig. 1** *Enhanced Workflow Architecture*

The proposed workflow architecture comprises several key stages designed to support high-performance real-time fraud detection and long-term analytical insight generation.

Data Ingestion:

Transaction data is sourced from banking APIs and third-party financial platforms, generating high-throughput streams of real-time data. These transactions are leveraged for both immediate money laundering detection and as training data for future model iterations. Apache Kafka serves as the primary ingestion layer, facilitating low-latency data streaming. Ingested data is temporarily stored in Amazon S3, capitalizing on its scalability, high availability, and support for unstructured formats to accommodate traffic surges and varied data types.

Data Processing:

Apache Spark is the central engine for data processing in both real-time and batch contexts, drawing input directly from Amazon S3.
- *Real-time Processing:* Spark Streaming processes transaction records in micro-batches to extract features relevant for rule-based and machine learning–based anomaly detection. Detected anomalies are forwarded to Apache Flink, which delivers low-latency alerts to downstream systems for immediate action. Flink's efficiency and responsiveness make it particularly well-suited for transactional stream analytics.
- *Batch Processing:* Transaction data accumulated over time is processed periodically for in-depth analytics and model retraining. This stage has less stringent latency requirements, allowing Spark to perform advanced computations and feature engineering, producing curated datasets for future ML model updates.

Feature Engineering and ML Model Development:

During batch processing, Spark computes engineered features such as transaction frequency, fraud ratios, and customer behavior metrics. These features are persisted in a structured backend database for use during inference. MLflow manages the machine learning lifecycle, monitoring metrics such as model accuracy and data drift. If deviations from defined thresholds are detected, retraining is automatically triggered, and updated models are deployed using the latest data retrieved from Amazon RDS.

Data Storage:

Amazon S3 is employed for the short-term storage of high-velocity, unstructured transaction data. For long-term storage and structured data management, Amazon RDS is used. This choice is guided by the stability of post-processed data schemas, RDS's cost-effectiveness, and its compatibility with analytical tools. As such, structured features and historical transaction records are stored in RDS, supporting scalable and sustainable data management.

Visualization and Reporting:

To support data-driven decision-making, the system integrates business intelligence tools such as Tableau and Power BI. These tools connect seamlessly to backend databases and offer robust visualization capabilities to present laundering trends, fraud risk scores, and historical metrics via dashboards. Additionally, Python libraries such as Matplotlib and Seaborn are utilized for statistical visualization and feature trend analysis, improving interpretability for analysts and stakeholders.

IV. MACHINE LEARNING MODELING AND RESULTS

This section outlines the data exploration, preprocessing steps, and machine learning techniques applied to identify patterns in fraudulent transactions and classify each transaction as either laundering or non-laundering. Python is used as the primary language for data preparation and model training, leveraging libraries such as *pandas*, *scikit-learn*, and *seaborn*.v

A. DATA PREPROCESSING

Initial data exploration and transformation were conducted to ensure the dataset was suitable for binary classification modeling. The preprocessing pipeline includes the following steps:
1. Data Loading and Feature Selection:
   Raw transaction data was ingested using *pandas*, and key features were extracted, including payment_currency, received_currency, sender_bank_location, receiver_bank_location, and payment_type. These features formed the basis of the final modeling dataset.

2. Data Splitting:
   The dataset was partitioned using *scikit-learn* into training, validation, and testing sets. This ensures robust model evaluation:
   - Training set: used to fit model parameters.

- Validation set: used to tune hyperparameters and compare model performance.
- Test set: used for final evaluation.

3. Handling Identifiers, Missing Values, and Time Variables:
Exploratory analysis showed no missing values or undocumented features. Identifier fields were dropped from all subsets, and time-related variables were excluded from this experiment for simplification.

4. Categorical Encoding:
Since all selected features are categorical, encoding techniques were applied to convert them into numerical formats suitable for model input.

5. Feature and Target Definition:
The independent variables (features) consist of the selected categorical fields. The dependent variable is the binary indicator is_laundering.

6. Handling Class Imbalance:
A significant class imbalance was observed, particularly in the payment_currency field, where over 90% of records involved UK pounds. Oversampling techniques were employed to address this imbalance and prevent the model from being biased toward the majority class.

*Figure 2* illustrates the imbalance in is_laundering across different payment currencies.

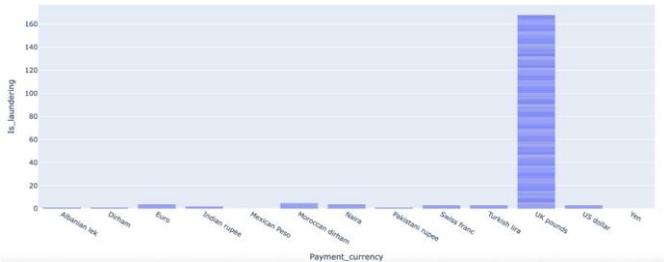

*Fig. 2 Is_ laundering by Payment_Currency*

B. Dataset Statistics

The final dataset subsets used for training and evaluation are as follows:
- Training data: 5,702,911 × 70
- Validation data: 1,900,970 × 70
- Test data: 1,900,971 × 70

C. Exploratory Analysis and Feature Insights

1. Feature Correlation:
A correlation heatmap as shown in figure 3 reveals relationships among features, indicating the need for additional feature engineering before model fitting.
2. Payment Type Distribution:
Analysis showed that laundering activity is disproportionately associated with certain payment types. As shown in *Table 1*, methods such as Cross-border transfers, Cash Withdrawals, and Cash Deposits exhibit higher fraud rates compared to card-based payments.

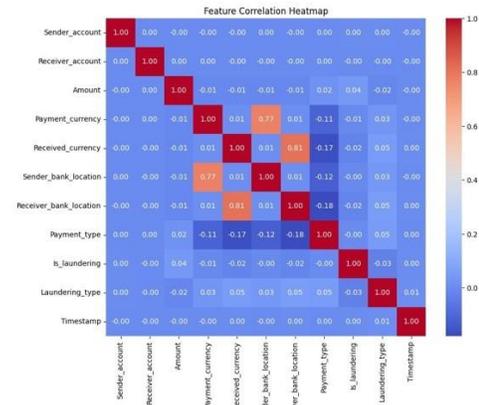

*Fig. 3 Correlation Heatmap*

*Table 1 Payment Type Distribution*

| Payment Type | Count | Fraudulent | % |
|---|---|---|---|
| Credit Card | 2,012,909 | 1136 | 0.06% |
| Debit Card | 2,012,103 | 1124 | 0.06% |
| Cheque | 2,011,419 | 1087 | 0.05% |
| ACH | 2,008,807 | 1159 | 0.06% |
| Cross-border | 933,931 | 2628 | 0.28% |
| Cash Withdrawal | 300,477 | 1334 | 0.44% |
| Cash Deposit | 225,206 | 1405 | 0.62% |

*Figures 3* and *4* visualize the alert distribution and fraudulent transaction count by payment type.

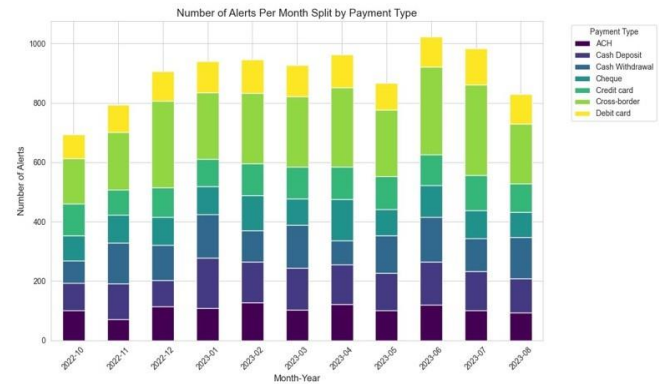

*Fig. 3 Number of Alerts Per Month Split by Payment Type*

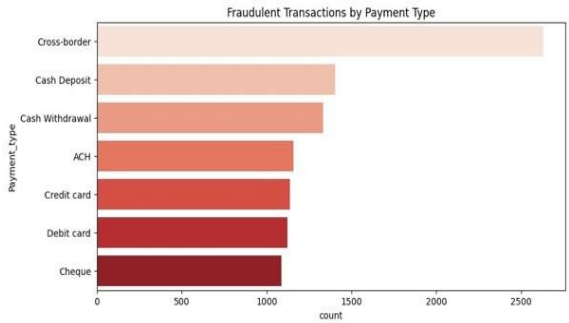

*Fig. 4 Fraudulent Transactions by Payment Type*

3. Seasonality Effects:
   *Figure 5* demonstrates that average fraudulent transaction amounts tend to peak around mid-year and year-end, aligning with major travel seasons. This suggests that time-of-year is a potential indicator of laundering activity.

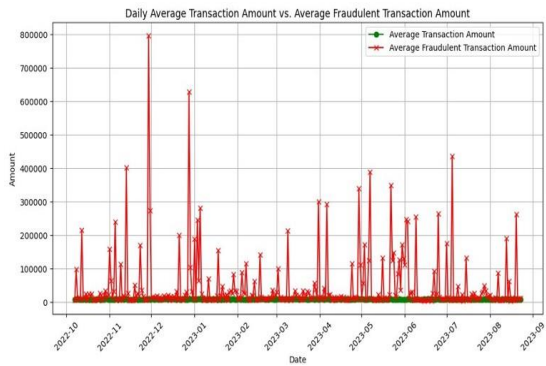

*Fig.5 Daily Average Transaction Amount vs. Average Fraudulent Transaction Amount*

D. Model Selection and Evaluation

Given the binary nature of the classification problem, the following models were implemented:
- Logistic Regression
- Decision Tree
- Random Forest

Model training and evaluation were conducted using *scikit-learn*, with results assessed using accuracy, F1-score, and confusion matrices.

1. Model Performance:
   Initial experiments yielded over 99% classification accuracy (*Figure 8*). However, such high accuracy may indicate overfitting due to class imbalance or data leakage. The confusion matrix (*Figure 6*) and F1-score comparison (*Figure 7*) support the need for improved feature selection and validation strategies.
2. Insights and Limitations:
   Despite promising accuracy, model generalizability remains a concern. The imbalance in class distribution necessitates further resampling, feature selection, and possibly ensemble approaches to enhance performance on minority classes.

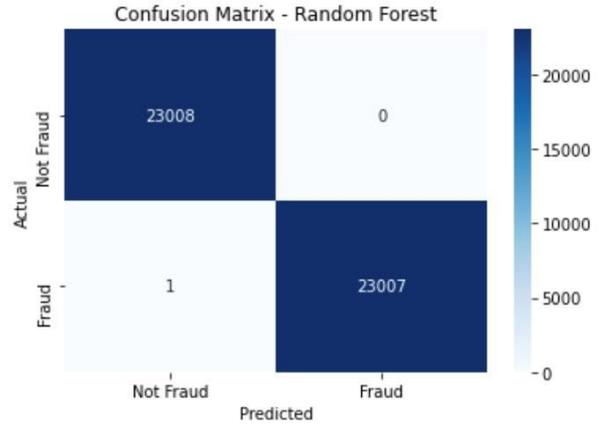

*Fig.6 Confusion Matrix*

```
                    Accuracy   F1-Score
Logistic Regression 0.998805   0.998806
Decision Tree       0.999891   0.999891
Random Forest       0.999978   0.999978
```

*Fig.7 F1-Score*

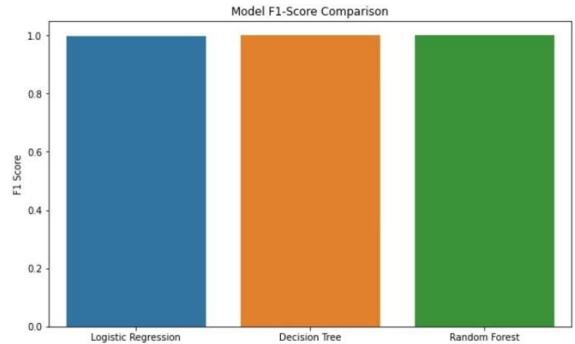

*Fig. 8 Model F1 Score Comparison*

## V. CONCLUSION

This project aimed to leverage Big Data technologies to extract actionable insights from financial transaction data and to enhance the detection of fraudulent activity. The results demonstrate that key features—such as transaction location, currency type, and temporal information—serve as strong indicators for classifying transactions as laundering or non-laundering.

Through the implementation of logistic regression, decision tree, and random forest models, it was shown that reliable predictive performance can be achieved using the selected features. The experimental results suggest that, despite class imbalance challenges, these models are capable of identifying laundering behavior with high accuracy when combined with appropriate data preprocessing and feature engineering techniques.

In parallel, the integration of Big Data tools—including Amazon S3, Amazon RDS, Apache Kafka, Apache Spark, and MLflow—provided practical insights into building scalable, real-time processing pipelines suitable for high-volume transaction data. These tools were explored not only for model training and deployment but also for end-to-end system design supporting ingestion, storage, processing, and visualization.

In conclusion, the findings of this study highlight the significant potential of Big Data technologies in empowering financial institutions to detect and mitigate fraudulent transactions more efficiently and at scale. Future work will explore the incorporation of temporal features, ensemble learning techniques, and real-world deployment scenarios for enhanced generalization and operational robustness.